\documentclass{appolb}
\usepackage{graphicx}
\usepackage{amssymb}
\usepackage{color}

\begin{document}
\title{Exploring nuclear exotica at the limits%
\thanks{Presented at XXVI Nuclear Physics Workshop 
2019 "Key problems of Nuclear Physics", 24-29 September 2019, Kazimierz
Dolny, Poland
}%
}
\author{A.\ V.\ Afanasjev, S.\ E.\ Agbemava and A.\ Taninah
\address{Department of Physics and Astronomy, Mississippi
State University, MS 39762}
}
\maketitle
\begin{abstract}
The study of nuclear limits has been performed and new physical mechanisms
and exotic shapes allowing the extension of nuclear landscape beyond the 
commonly accepted boundaries have been established. The transition from
ellipsoidal to toroidal shapes plays a critical role in potential extension of nuclear
landscape to hyperheavy nuclei. Rotational excitations leading  to the birth of
particle (proton or neutron) bound rotational bands provide a mechanism for
an extension of nuclear landscape beyond spin zero proton and neutron 
drip lines. 
\end{abstract}
\PACS{PACS numbers come here}
  
\section{Introduction}

The studies of the nuclei at the limits are guided by human curiosity,
by the need to understand new physical mechanisms governing nuclear
systems in these extreme conditions and by the demand for nuclear
input in nuclear astrophysics. There is the set of the questions related to
the physics at the limits listed below which looks deceivably simple but 
extremely difficult to answer. These questions are:  What are the limits 
of the existence of nuclei? What are the highest proton number $Z$ at 
which the nuclear landscape and periodic table of chemical elements 
cease to exist? What are the positions of proton and neutron drip 
lines? Are there some physical mechanisms which allow to extend 
the nuclear landscape beyond spin-zero limits? What type of nuclear 
shapes dominate these extremes of nuclear landscape?

\begin{figure}[htb]
\centerline{%
\includegraphics[width=5.3cm,angle=-90]{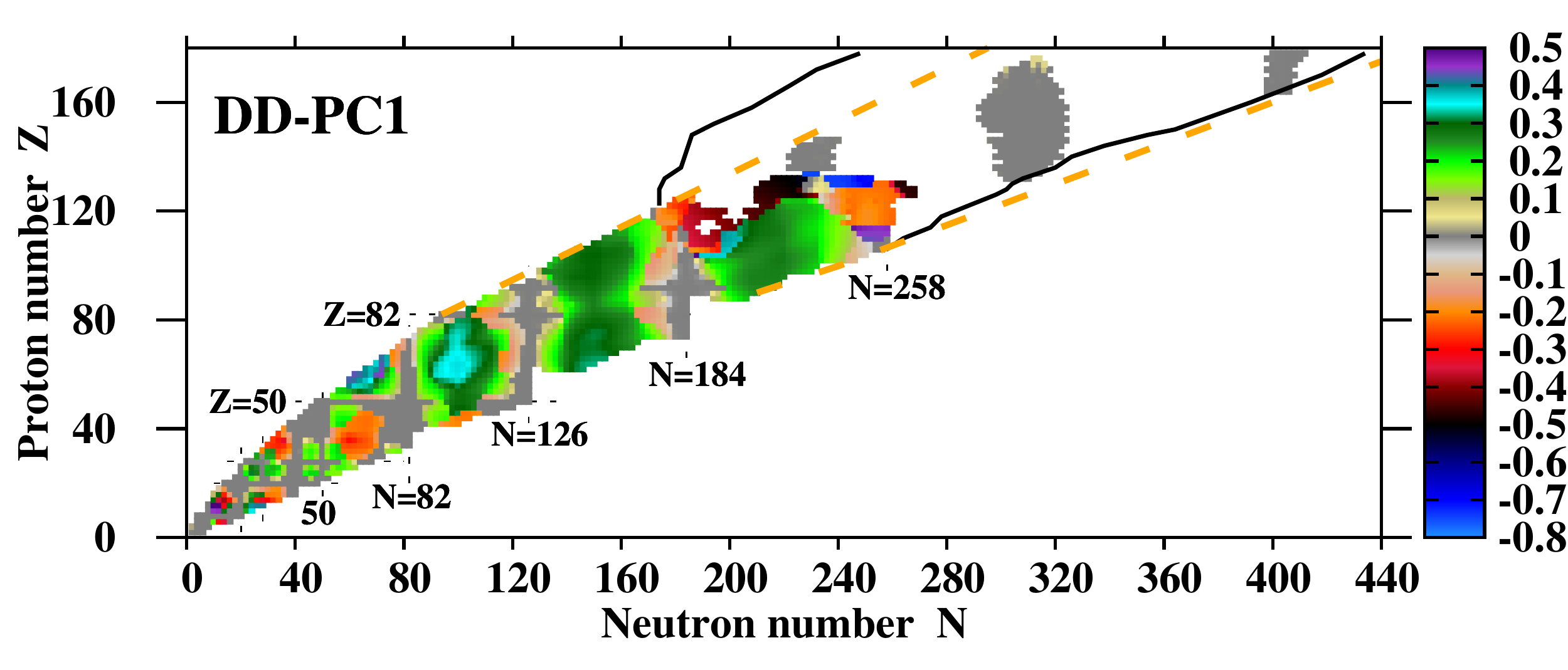}
}
\caption{The distribution of ellipsoidal and toroidal shapes in nuclear 
landscape.  The nuclei with ellipsoidal shapes are shown by the squares 
the color of which indicates the equilibrium quadrupole deformation $\beta_2$ 
(see colormap).  Note that ellipsoidal shapes with the heights of fission barriers 
smaller than 2.0 MeV are considered as unstable (see the discussion in Sect. III
of Ref.\ \cite{AAR.12}.)  Two-proton and two-neutron drip lines for toroidal nuclei are 
shown by solid black lines. The white region between them (as well as the 
islands inside this region shown in gray) corresponds to the nuclei  which have 
toroidal shapes in the lowest in energy minimum for axial symmetry (LEMAS).
The islands of relatively stable spherical hyperheavy nuclei in the  $Z>130$ nuclei, 
shown in light grey color, correspond  to the solutions which are excited in energy 
with respect of the LEMAS corresponding to toroidal shapes. 
Note that in the same nucleus two-neutron drip lines for spherical and toroidal 
shapes are different. This is the reason why some islands of stability of spherical 
hyperheavy nuclei extend beyond the two-neutron drip line for toroidal shapes. 
The extrapolations of the two-proton and two-neutron drip lines for ellipsoidal 
shapes, defined from their general trends seen in the $Z<120$ nuclei, are
displayed by thick orange dashed lines.}
\label{Landscape}
\end{figure}

Over recent years our group has undertaken a systematic efforts
in the studies of these questions within the framework of covariant density
functional theory (CDFT) \cite{VALR.05}. The analysis of Refs.\ 
\cite{AARR.13,AARR.14,AA.16} performed with three major classes of 
covariant energy density functionals  (CEDF) has allowed to evaluate the 
global performance of these functionals in the description of the ground state 
properties of even-even nuclei. It addition, it permitted to estimate systematic 
uncertainties \cite{DNR.14} in their 
description and the propagation of these uncertainties on approaching nuclear 
limits for the $Z \leq 106$ nuclei.  Moreover,  such an analysis has allowed to 
estimate theoretical uncertainties in the predictions
of the two-proton and two-neutron drip lines for the $Z \leq 120$ nuclei
(see Refs.\ \cite{AARR.13,AARR.14,AARR.15}) and compare them with 
those obtained in non-relativistic theories (see Refs.\ \cite{Eet.12,AARR.14}). 
Note that in the  CDFT  statistical uncertainties emerging from the details 
of the fitting  protocol are significantly smaller than systematic ones originating 
from  the choice of the form of the functional \cite{AAT.19}. This is contrary to 
the case of non-relativistic Skyrme energy density functionals in which these 
two types of uncertainties are comparable at nuclear limits 
\cite{Eet.12,GDKTT.13}.

 These results formed the basis for a subsequent analysis of the extension
of nuclear  landscape to hyperheavy nuclei \cite{AAG.18,AATG.19}, the study 
of the properties of toroidal hyperheavy nuclei \cite{AAG.18,AATG.19} and the 
discovery of new mechanism of the extension of nuclear landscape beyond 
spin-zero limit by means of rotational excitations \cite{AIR.19}. In the present
paper, they will be briefly reviewed and, in addition, will be supplemented by 
new results.

The paper is organized as follows. Sec.\ \ref{Toroidal} is dedicated to the discussion  of toroidal 
shapes in hyperheavy nuclei. The impact of rotational excitations on the 
boundaries of nuclear landscape and underlying physical mechanism
are considered in Sect.\ \ref{Rotational}.  Finally, the summary is presented in 
Sect.\ \ref{Summary}.

\section{The dominance of toroidal shapes in hyperheavy $Z\geq 126$ nuclei}
\label{Toroidal}

\begin{figure}[htb]
\centerline{%
\includegraphics[width=12.5cm]{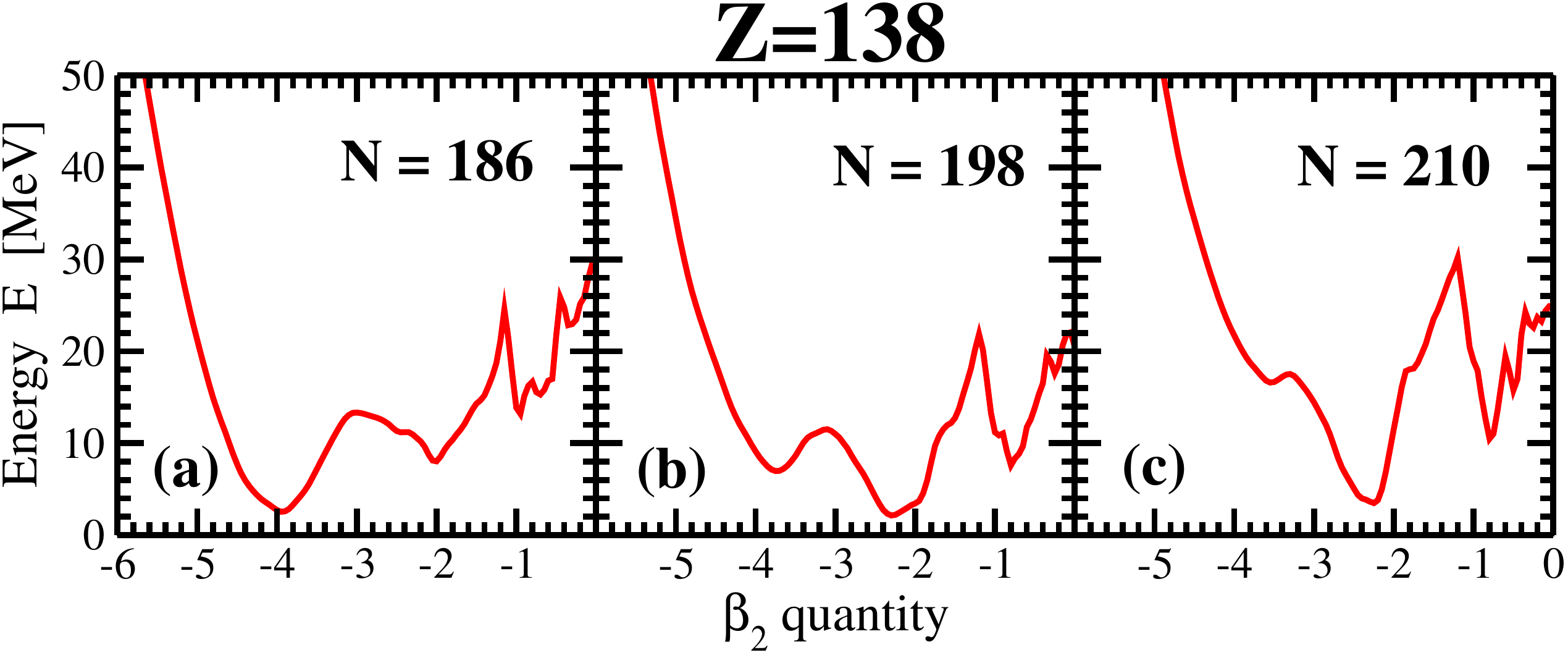}
}
\caption{Deformation energy curves of selected $Z=138$ hyperheavy nuclei 
obtained in axial RHB calculations with the DD-PC1 functional.
}
\label{PEC-Z=138}
\end{figure}

   The state-of-the-art view on the nuclear landscape is presented in Fig.\ 
\ref{Landscape}. It is born out of systematic axial relativistic Hartree-Bogoluibov 
(RHB) calculations of Refs.\ \cite{AARR.14,AAG.18,AATG.19} supplemented by 
triaxial RHB calculations for the ground states and fission barriers 
of selected set of the nuclei \cite{AAG.18,AATG.19}. These calculations are
based on the DD-PC1 functional \cite{DD-PC1} which is one of the best 
CEDFs \cite{AARR.14}.  For the $Z<100$ nuclei
we see the classical nuclear structure with pronounced spherical shell gaps
at particle numbers 8, 20, 28, 50, 82 (and $N=126$) leading to the bands 
(shown by gray color) of spherical nuclei in the nuclear chart along the vertical
and horizontal lines with these particle numbers. In addition, the traditional picture 
of the transition from spherical shapes to prolate ones, then to oblate ones and finally 
to spherical shapes on moving from one spherical shell closure to another one
is seen for these nuclei.

\begin{figure}[htb]
\centerline{
\includegraphics[width=11.5cm]{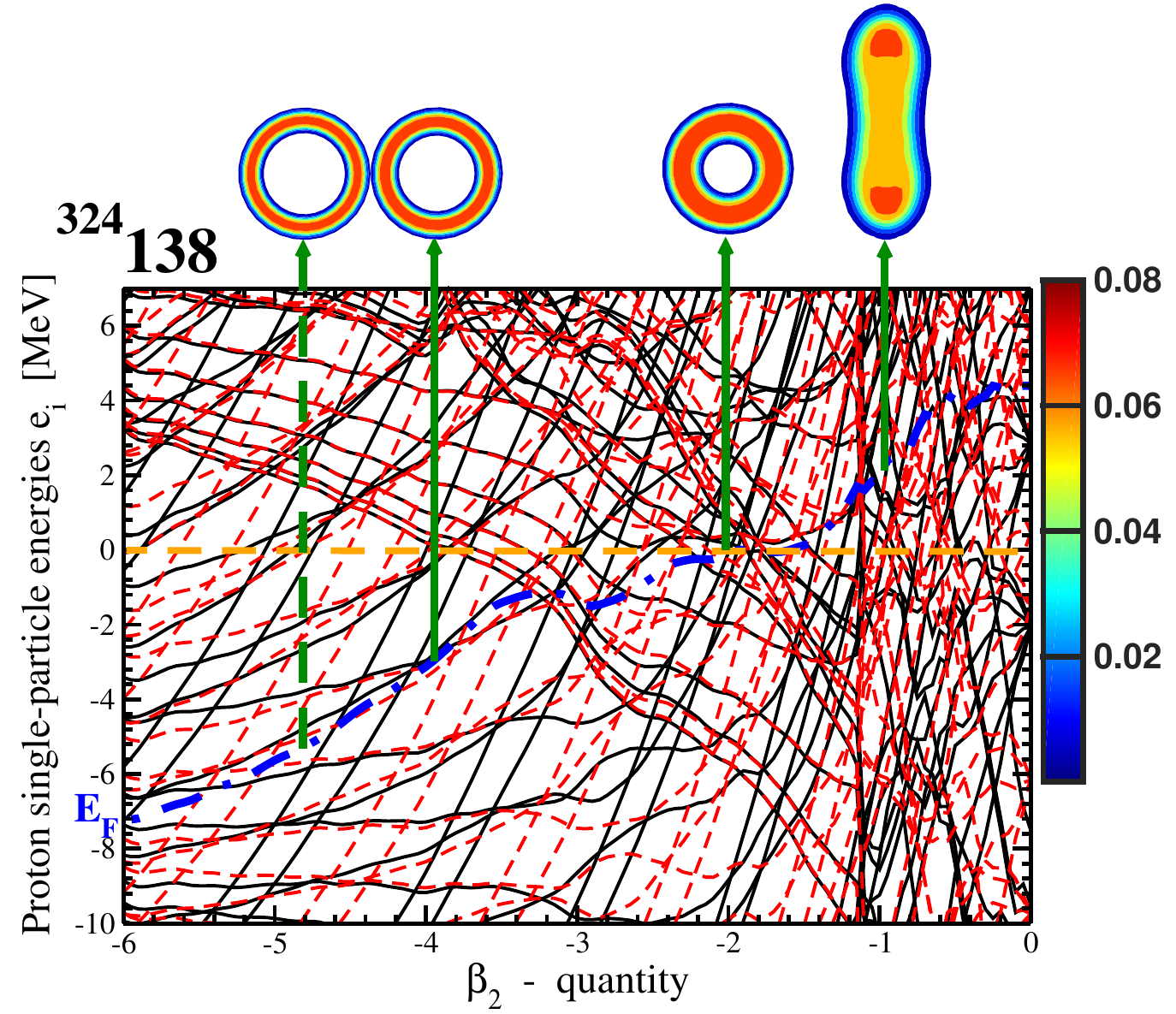}}
\caption{Proton single-particle energies for the lowest in total energy 
solution in the nucleus $^{324}138$ calculated as a function
of the $\beta_2$ quantity. Solid and dashed lines are used for 
positive- and negative-parity states, respectively.
Proton density distributions, corresponding to local
minima shown in Fig.\ \ref{PEC-Z=138}(a)  as well as to the solution 
with $\beta_2=-4.8$, are displayed at the top of the figure. They are 
shown in the plane containing toroid for toroidal shapes and in 
the  plane containing the axis of symmetry for biconcave disk.
The density colormap starts at $\rho_n=0.005$ fm$^{-3}$ and shows 
the densities in fm$^{-3}$. Blue dot-dashed and orange dashed lines 
are used for the Fermi level ${\rm E}_{\rm F}$  and zero energy, 
respectively. }
\label{Stable-vs-unstable}
\end{figure}

   However, these features are in general gone for superheavy ($Z=100-124$)
nuclei. In this region, normal- and superdeformed oblate shapes become
dominant for high $Z$ nuclei (see Fig.\ \ref{Landscape}). Further increase of
proton number triggers the transition to toroidal shapes\footnote{Many features
of toroidal shapes in nuclear physics are discussed in Ref.\cite{W.73} but 
these discussions do not extend to hyperheavy nuclei. In recent years, there
is an increased interest to toroidal high-spin isomers (see, for example, 
Refs.\ \cite{SW.14,SWK.17,IMMI.14}).}.  This transition is driven
by Coulomb repulsion: for high-$Z$ systems the Coulomb energy is significantly 
larger for the ellipsoidal shapes than for toroidal ones (see discussion 
in Sec. XII of Ref.\ \cite{AATG.19}).   As a consequence, the toroidal shapes,
corresponding to large negative values of the $\beta_2$ quantity\footnote{The  
$\beta_2$ and $\gamma$ quantities are extracted
from respective quadrupole moments:
\begin{eqnarray}
Q_{20} &=& \int d^3r \rho({\vec r})\,(2z^2-x^2-y^2),\\
Q_{22} &=& \int d^3r \rho({\vec r})\,(x^2-y^2),
\end{eqnarray}
via
\begin{eqnarray}
\beta_2 &=&  \sqrt{\frac{5}{16\pi}} \frac{4\pi}{3 A R_0^2} \sqrt{ Q_{20}^2 + 2Q_{22}^2}
\\
\gamma&=& \arctan{\sqrt{2} \frac{Q_{22}}{Q_{20}}}
\end{eqnarray}
where $R_0=1.2 A^{1/3}$. Note that $Q_{22}=0$ and $\gamma=0$ in
axially symmetric RHB calculations. The $\beta_2$ and 
$\gamma$ values have a standard meaning of the deformations only
for 
ellipsoid-like density distributions with $|\beta_2| \lesssim 1.0$.
For toroidal shapes ($\beta_2 \lesssim -1.2) $ they should be treated 
as dimensionless and particle normalized measures of  the $Q_{20}$ 
and $Q_{22}$ moments of the density distributions. The $\beta_2$
quantity defines the radius $R$ of the toroid and its tube radius $d$:
with increasing absolute value of the $\beta_2$ quantity $R$ increases
and $d$ decreases. The $\gamma$-quantity defines the deviation of the 
density distribution from symmetric ($\gamma=60^{\circ}$) toroid.}
become energetically favored as compared with ellipsoidal ones in the nuclei 
with high-$Z$ values. This is illustrated in Fig.\ \ref{PEC-Z=138} where the 
competition  of three local minima with $\beta_2 \sim -0.8$,  $\beta_2 \sim -2.3$, and 
$\beta_2 \sim -3.8$ are clearly seen. The first minimum corresponds to 
biconcave disk (oblate) shapes, while other two to toroidal shapes. 
Dependent on the combination of proton and neutron numbers, one of these
minimum becomes LEMAS. In lower $Z$ nuclei, the biconcave disk shape
corresponds to LEMAS (see Fig.\ \ref{Landscape}).  The LEMAS in the nuclei in 
the vicinity of the $Z\sim 136, N\sim 210$ corresponds to toroidal shapes with 
$\beta_2 \sim -2.3$  [Fig.\ \ref{PEC-Z=138}(c) and Ref.\ \cite{AA.20}). Going away from 
this region favors toroidal shapes with  substantially larger (in absolute value)
$\beta_2$ quantities [Fig.\ \ref{PEC-Z=138}(a) and Ref.\ \cite{AA.20}]. Typical 
density distributions  corresponding to these shapes are shown at the top of 
Fig.\ \ref{Stable-vs-unstable}.

Fig.\ \ref{Landscape} compares the extrapolations of the two-neutron and 
two-proton drip lines for ellipsoidal shapes 
with those obtained for toroidal nuclei.
Note that these extrapolations are quite close to respective drip lines 
obtained for the islands of spherical hyperheavy nuclei. The calculated 
two-neutron drip line for toroidal shapes is close to the extrapolation 
of this line for ellipsoidal shapes. On the contrary, the situation is complely
different on the proton-rich side of the nuclear landscape: the transition to toroidal
shapes in the $Z > 120$ nuclei creates a substantial expansion [the area
between black solid and orange dashed lines in Fig.\ \ref{Landscape}] of the nuclear 
landscape. The reason for that is clearly seen in Fig.\ \ref{Stable-vs-unstable} on the
example of the $^{324}138$ nucleus.  The Fermi level E$_{\rm F}$ for ellipsoidal-like 
shapes  is located at  positive energies and thus such shapes are unstable against 
proton emission. On the contrary, the transition  to toroidal shapes drastically modifies 
the  underlying single-particle  structure and as a consequence  lowers the energy 
of the Fermi level with increasing absolute value of the $\beta_2$ quantity.  As 
a consequence, $E_{\rm F}\approx -3.5$ MeV for LEMAS with $\beta_2\sim -4.0$ 
in this nucleus and thus  this state is particle bound. 

\begin{figure*}[htb]
\centering
\includegraphics[angle=0,width=6.2cm]{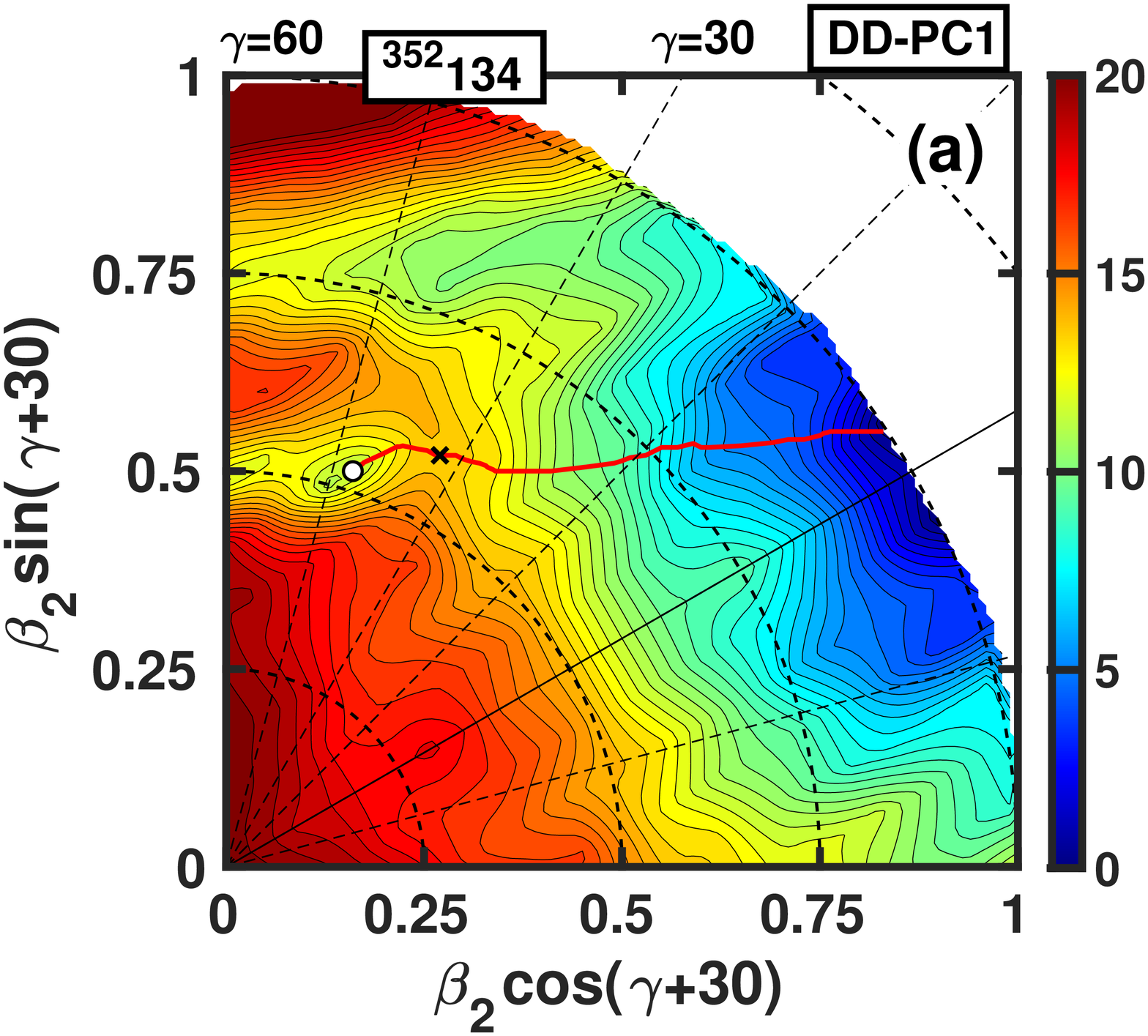}
\includegraphics[angle=0,width=6.2cm]{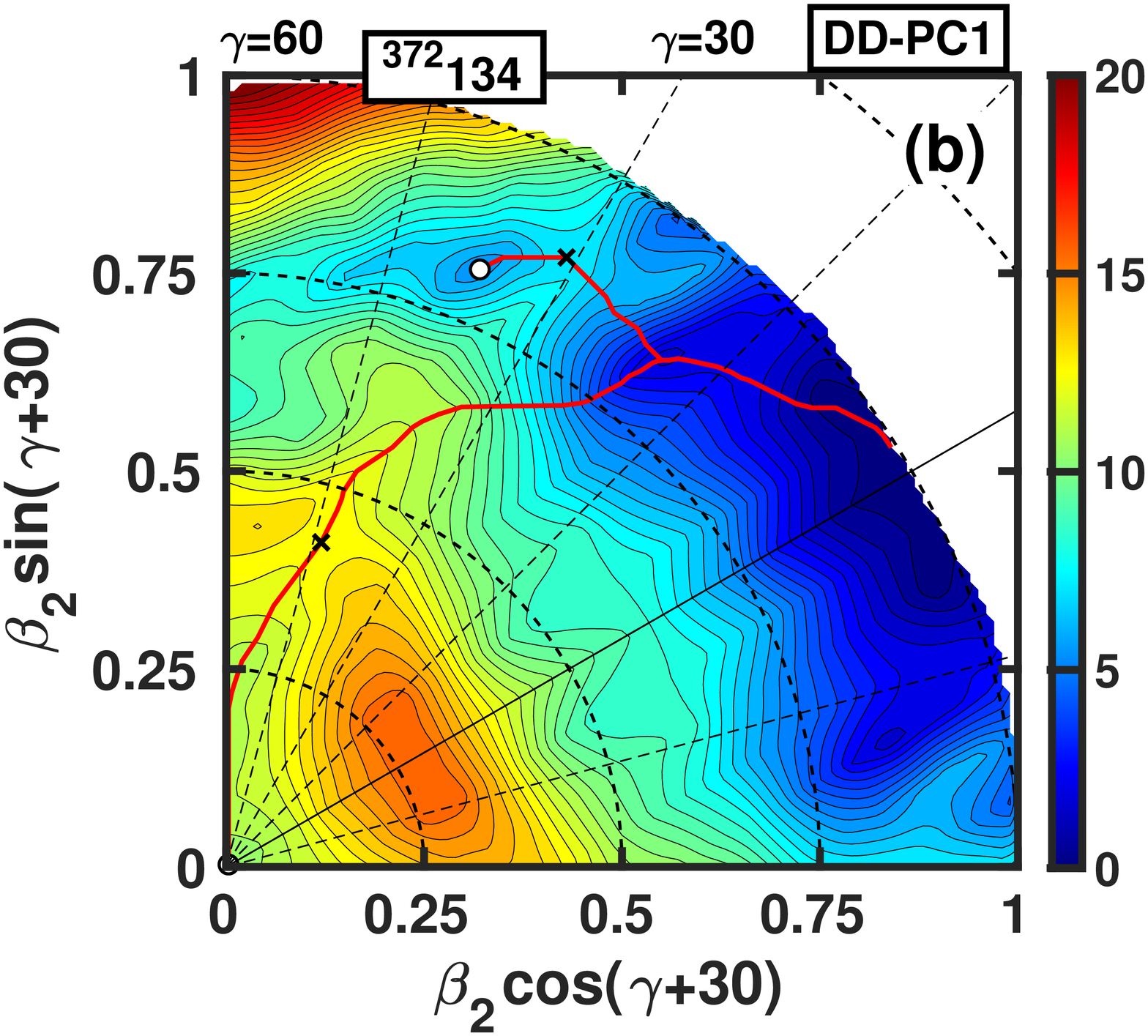}
\includegraphics[angle=0,width=6.2cm]{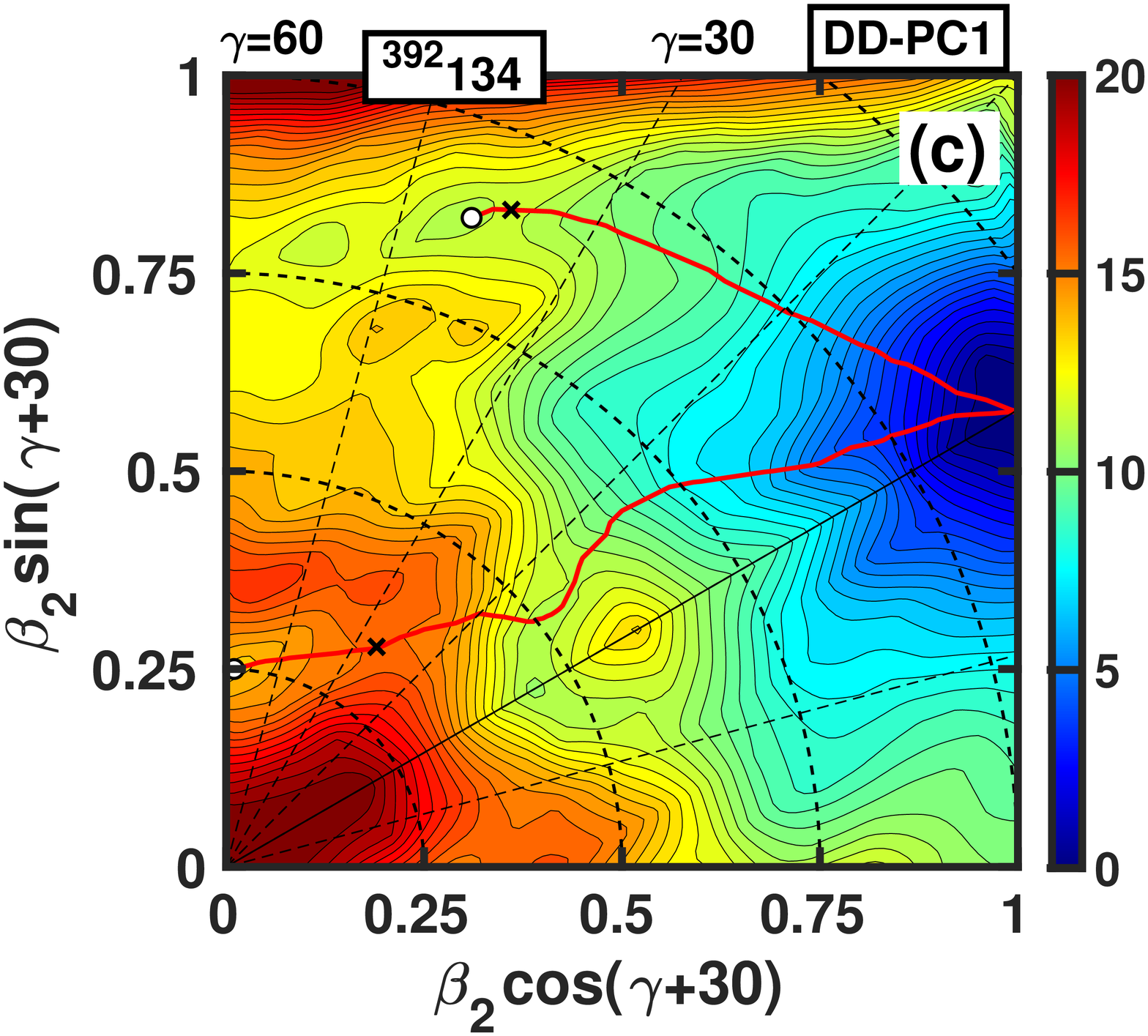}
\includegraphics[angle=0,width=6.2cm]{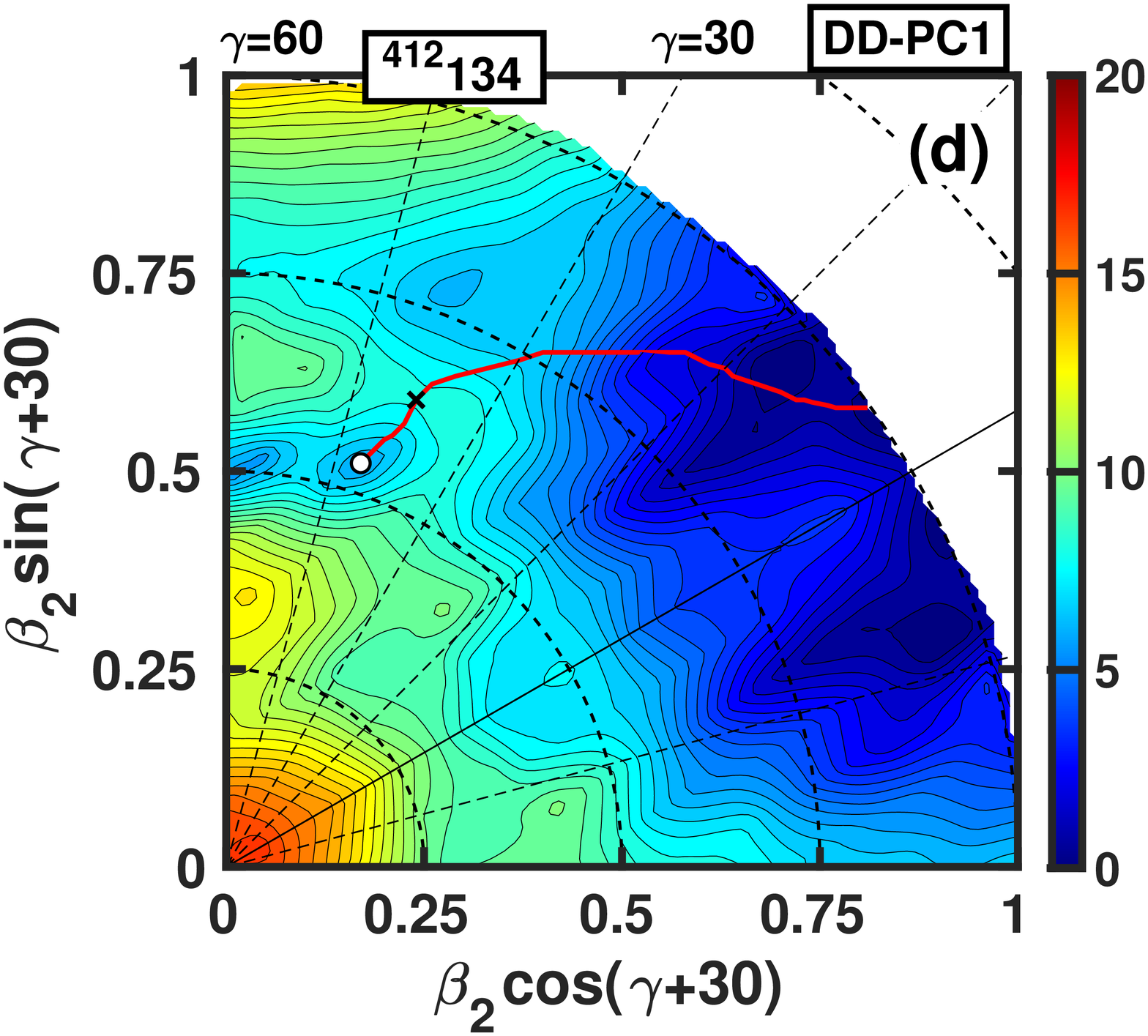}
\caption{Potential energy surfaces (PES) of indicated nuclei obtained in
the RHB calculations. The energy difference between two neighboring 
equipotential lines is equal to 0.5 MeV. The red lines show static 
fission paths from respective minima. Open white circles show the
global (and local) minimum(a). Black crosses indicate the saddle 
points on these fission paths. The colormap shows the excitation
energies (in MeV) with respect to the energy of the deformation 
point with largest binding energy. The panel with the results for 
$^{392}$134 nucleus is taken from Ref.\ \cite{AAG.18}.}
\label{triaxial}
\end{figure*}

  The triaxiality plays a critical role in the demise of ellipsoidal shapes and 
the emergence of toroidal shapes as a major player in hyperheavy nuclei.  This
is because the impact of triaxiality on fission barriers gets much more pronounced 
in the nuclei with ground state oblate shapes and it generally increases with the 
rise of their oblate deformation \cite{AAG.18,AATG.19}.  Not only the fission through the $\gamma$-plane 
gets more energetically favored, but also the fission path through $\gamma$-plane 
becomes much shorter than the one through the $\gamma=0^{\circ}$ axis.  These 
features are illustrated in Fig.\ \ref{triaxial} on few $Z=134$ nuclei distributed 
equidistantly in neutron number along the isotopic chain. Let us consider the 
$^{392}$134 nucleus as an example. In axial RHB calculations, its ground state 
and excited (at excitation energy of 2.69 MeV) minima are located at 
$\beta_2=-0.79$ and $\beta_2=-0.23$, respectively. The fission 
barriers for these two minima are 10.24 and 7.55 MeV, respectively.  
The inclusion of the triaxiality reduces these fission barriers down to 
0.56 and 2.08 MeV making ellipsoidal shapes unstable with respect
of fission. Similar effects are also seen in the $^{352}$134, $^{372}$134,
$^{412}$134 (see Fig.\ \ref{triaxial}) and $^{432}$134 (see Ref.\ \cite{AAG.18}) 
nuclei. Note that similar trend of substantial reduction of fission barriers for 
ellipsoidal shapes in hyperheavy nuclei due to triaxiality is seen also in
microscopic+macroscopic  and Skyrme DFT calculations  \cite{BS.13}.

\begin{figure}[htb]
\centerline{%
\includegraphics[width=5.5cm]{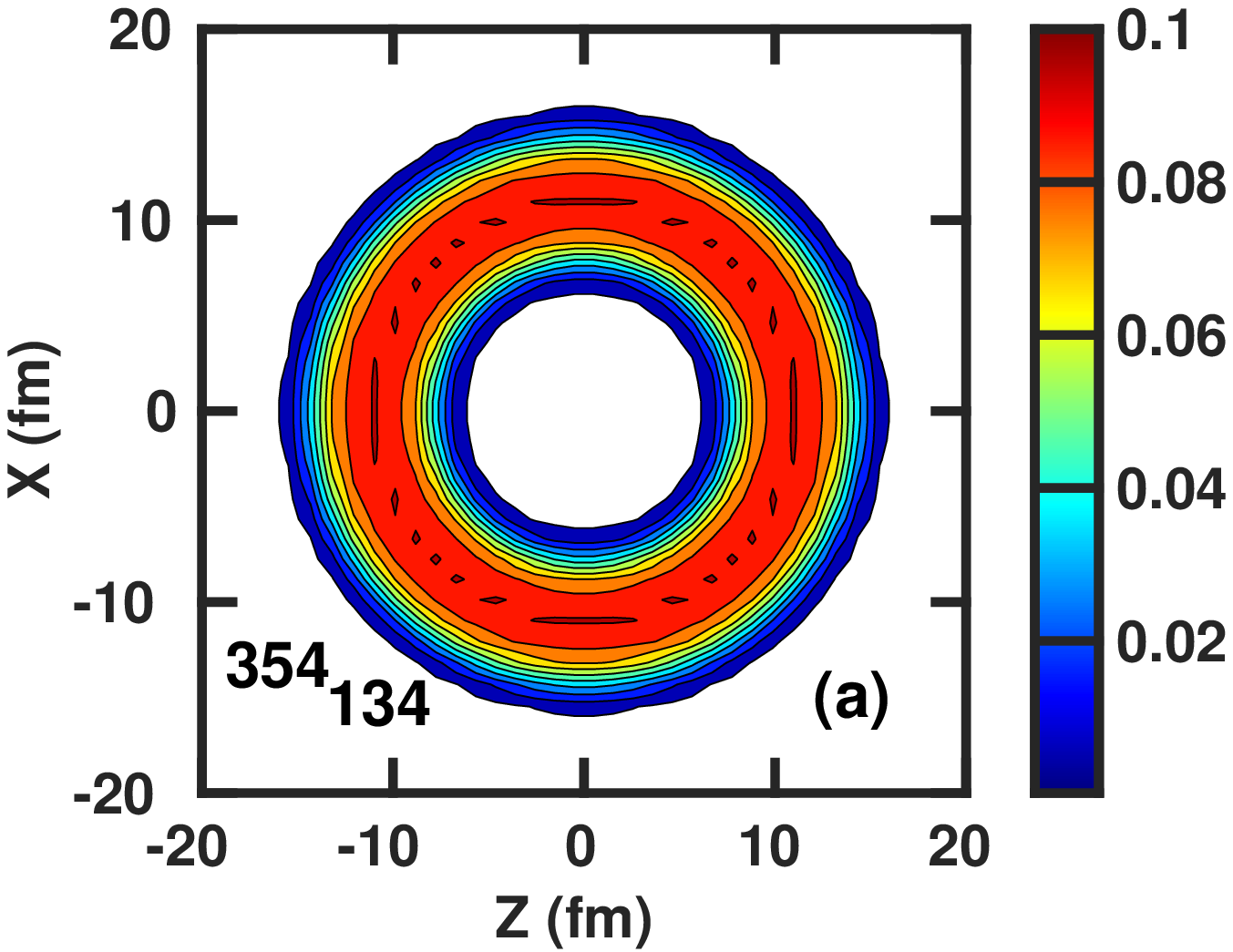}
\includegraphics[width=5.5cm]{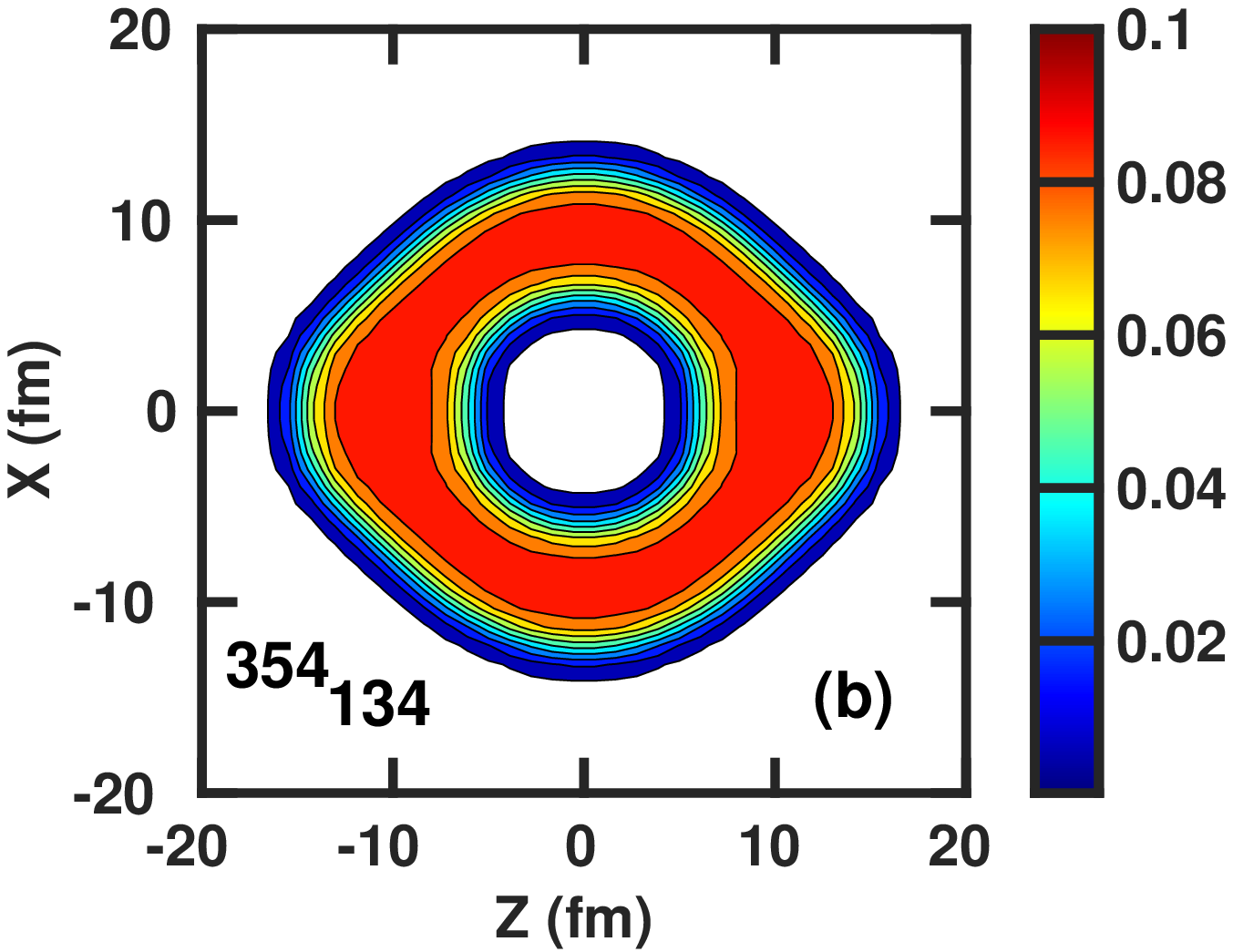}
}
\caption{Neutron density distributions  of the lowest stable toroidal configuration 
              with $\beta_4 \sim 1.5$  in the $^{354}$134 nucleus corresponding to its
               minimum with $\beta_2 \sim 2.3, \gamma=60^{\circ}$ [left panel] 
               and the saddle with $\beta_2 \sim 2.0, \gamma \approx 35^{\circ}$               
               [right panel].  The density colormap starts at $\rho_n=0.005$ fm$^{-3}$ 
               and shows the densities in fm$^{-3}$.     
}
\label{Density}
\end{figure}

   The investigation of potential stability of toroidal nuclei with respect of different 
types of distortions is one of important aspects of their studies.  The results of 
systematic axial RHB calculations of Refs.\ \cite{AAG.18,AATG.19} clearly indicate 
that toroidal nuclei are stable with respect of breathing deformation\footnote{Similar
results have also been obtained in the HFB calculations with the Gogny force but only
for two nuclei \cite{Warda.07}.}. Note that breathing 
deformation preserves the azimuthal symmetry of the torus  and it is defined by 
the radius of torus and the radius of its tube \cite{W.73}. 
Another type of the distortions of toroidal shapes is with respect of so-called 
sausage deformations; they make a torus thicker in one section(s) and 
thinner in another section(s) \cite{W.73}. Such a distortion, corresponding
to sausage deformation of order $\lambda=2$, is examplified in Fig.\ \ref{Density} as 
the transition from the density distribution shown in panel (a) to that 
shown in panel (b). The study of such deformations requires triaxial RHB 
or RMF+BCS computer codes and could be potentially performed only for 
toroidal shapes in the $Z\sim 134, N \sim 210$ region, which are characterized by small 
radius of  torus and large radius of its tube, because of enormous requirements
on the basis size (see Sect. III in Ref.\ \cite{AATG.19}). Even then such calculations  
are extremely time-consuming. As a result,  the stability of toroidal hyperheavy 
nuclei with respect of non-axial distortions related to sausage deformations of 
order $\lambda=2$ has  been studied so far for only two nuclei, namely, 
$^{354}134$ and $^{348}138$ in Refs.\ \cite{AAG.18,AATG.19}. The saddle 
points of their fission barriers with respect of such distortions are located at 
4.4 and 8.54 MeV, respectively; this is indicative of their potential stability.
 
   The situation with odd order sausage deformations with $\lambda=1$ and 
3 is even more complicated because its resolution  requires the use of spatial 
symmetry unrestricted computer codes. However, numerical calculations with 
basis sizes following from the analysis of Sect. III of Ref.\ \cite{AATG.19} are 
impossible nowadays in such computer codes.

\section{Extension of nuclear landscape by means of rotational excitations}
\label{Rotational}

\begin{figure}[ht]
\centering
\includegraphics[width=8.5cm]{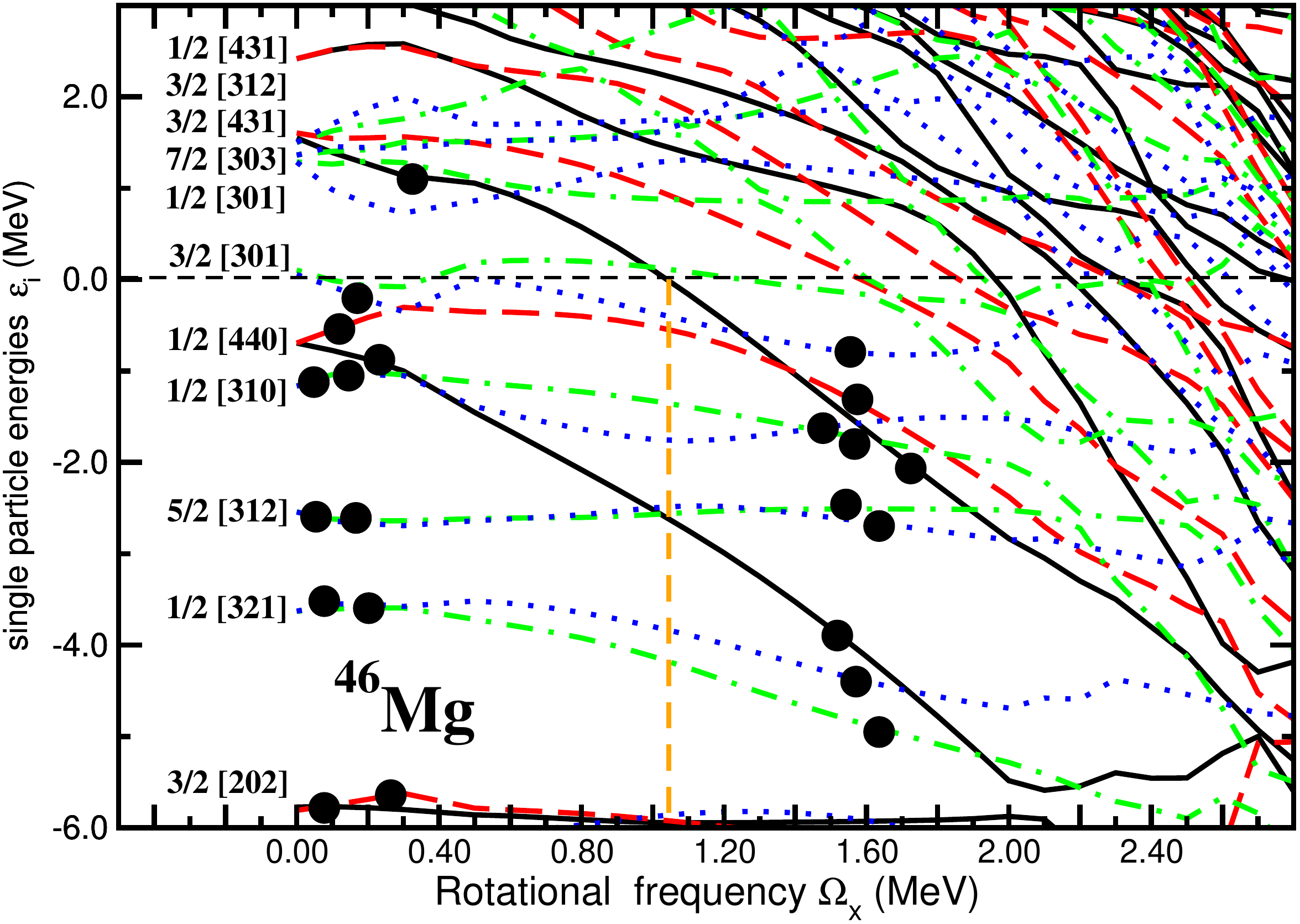}
\caption{Neutron single-particle energies (routhians) 
in the self-consistent rotating potential of $^{46}$Mg as a function of 
rotational frequency $\Omega_x$. They are given along the deformation 
path of the [3,0] configuration. Long-dashed  red, solid black, dot-dashed green,
and  dotted blue lines indicate
$(\pi = +, r = +i)$, $(\pi = +, r = -i)$,
$(\pi = -, r = +i)$, and $(\pi = -, r = -i)$ orbitals, respectively.
At $\Omega_x = 0.0$ MeV, 
the single-particle orbitals are labeled by the asymptotic quantum 
numbers $\Omega[Nn_z\Lambda]$ (Nilsson quantum numbers) of the dominant 
component of the wavefunction. Solid circles indicate occupied orbitals in
resonance and particle-bound parts of  respective configurations.  Vertical orange 
dashed line indicates the frequency at which the configuration becomes particle bound.
}
\label{routh-Mg46}
\end{figure}

   Two new mechanisms active in rotating nuclei located in the vicinity of 
neutron drip line have been discovered by us in Ref.\ \cite{AIR.19}. This 
investigation has been performed in the cranked relativistic mean field (CRMF) 
approach \cite{KR.90,TO-rot} without pairing; the neglect of pairing is a 
reasonable approximation for high spins of interests (see Ref.\ \cite{AIR.19}
for details). Strong Coriolis interaction acting on high-$j$ orbitals transforms 
particle-unbound (resonance)  nucleonic configurations into particle-bound ones 
with increasing angular momentum. The point of the transition manifests {\it the 
birth of particle-bound rotational bands.}  This mechanism is best illustrated on 
the example of the [3,0]\footnote{This configuration contains 3 neutrons in the
intruder $N=4$ orbitals and no protons in the intruder $N=3$ orbitals; see 
Ref.\ \cite{AIR.19} for more details. } configuration in $^{46}$Mg the neutron 
routhian diagram of which is shown in Fig.\ \ref{routh-Mg46}.  The $3/2[431](r=-i)$ 
orbital is the highest in energy occupied positive parity intruder orbital in this 
configuration.  In each (proton or neutron) subsystem, the energy of the highest 
occupied orbital corresponds  to the energy of the Fermi level in the calculations 
without pairing  \cite{NilRag-book}. Thus, the nucleonic configuration, the
energy of the last occupied neutron orbital of which is negative, is expected
to be particle bound. At rotational frequency $\Omega_x <1.03$ MeV, the $3/2[431](r=-i)$ 
orbital is particle unbound since its single-particle energy is positive (see Fig.\ \ref{routh-Mg46}). 
Thus, at low spin (up to $I\sim 16\hbar$ corresponding to $\Omega_x=1.03$ MeV), the 
rotational band built on this configuration can exist only as a band embedded  in 
particle  continuum (see Refs.\ \cite{GEF.13,FNJMP.16}) (further 'resonance band').
Above this frequency, its energy becomes negative and this orbital dives
(because of its high-$j$ content leading to strong Coriolis force) deeper
into nucleonic potential with increasing rotational frequency. As a consequence,
the [3,0] configuration becomes particle bound. Thus, respective rotational band
changes its character from  particle-unbound resonance band (at $I<16 \hbar$)
to particle-bound rotational band  (at $I>16\hbar$) with discrete rotational
states  of extremely narrow width.  Alternative possibility of the transition from
particle-bound to resonance rotational band ({\it the death of particle-bound
rotational bands}\footnote{The stability of aligned states in the $^{40}$Ca, $^{66}$Ge, 
$^{122}$Xe and $^{150}$Gd nuclei with increasing spin has been studied in Ref.\ 
\cite{CDZALANR.77}; above some spin value such states become particle unstable. 
Despite some similarities of our results and those of Ref.\ \cite{CDZALANR.77}, there
are important differences. We consider collective rotation of the nuclei,  while 
the regime of nuclear motion in aligned states of Ref.\ \cite{CDZALANR.77} 
corresponds to so-called "non-collective rotation" (see Refs.\ \cite{NilRag-book,PhysRep-SBT}). 
These aligned states in many cases represent non-collective terminating
states of rotational bands (or highest spin state within a rotational band) 
\cite{PhysRep-SBT}.}) with increasing spin also exists but it is less frequent
in the calculations.

\begin{figure}[ht]
\centering
\includegraphics[width=10.5cm]{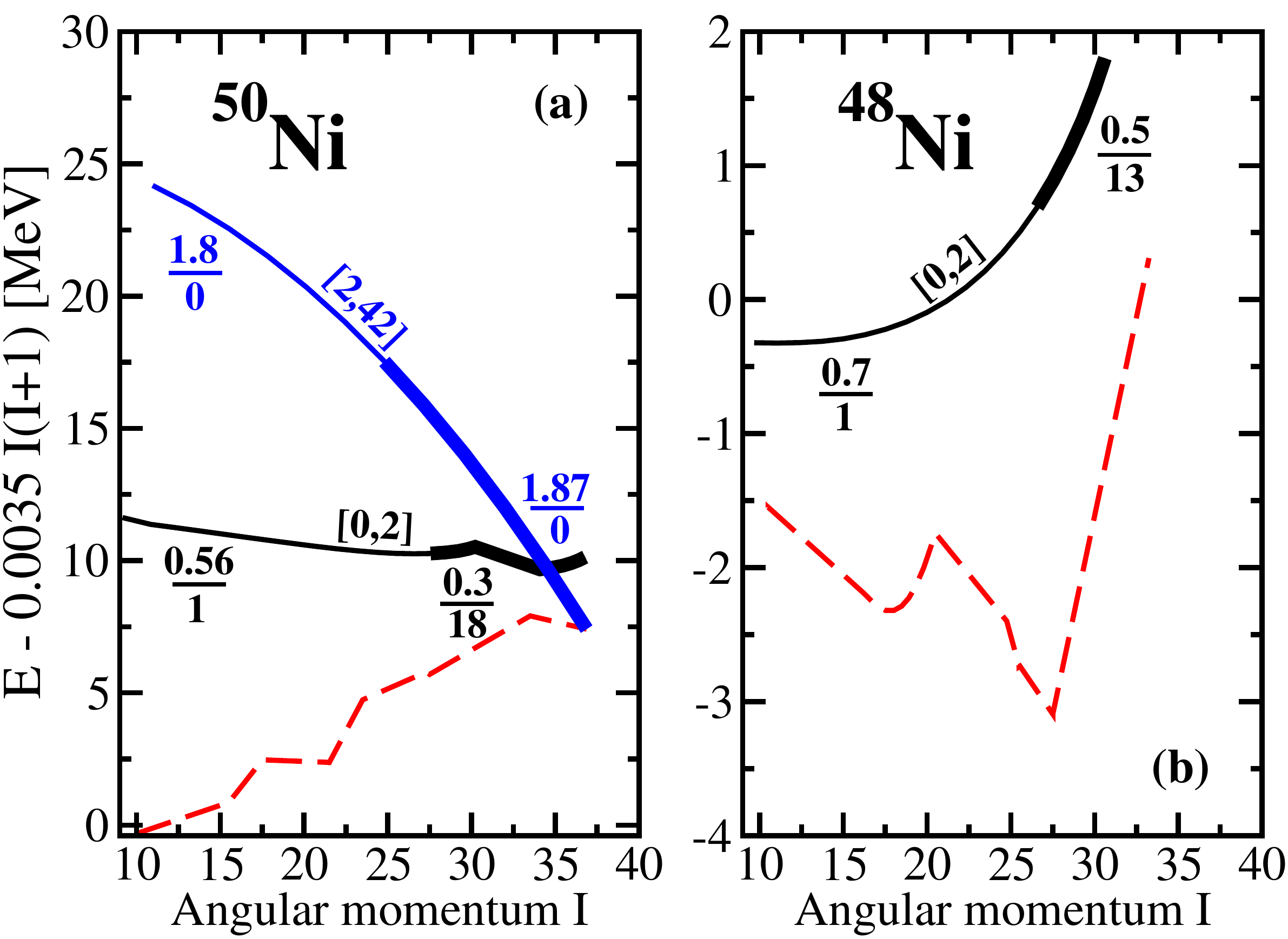}
\caption{Excitation energies of calculated configurations in $^{48,50}$Ni
relative to a rotating liquid drop reference $AI(I+1)$, with the inertia parameter 
$A=0.0035$. Thin and thick lines show proton emitting and proton bound 
parts of rotational bands, respectively. The yrast lines (as defined from 
approximately 15 calculated configurations) are shown by red dashed
lines.  Typical deformations of the bands  are shown in the format 
$\frac{\beta_2}{\gamma}$.  The configurations in proton-rich Ni isotopes are labeled by the shorthand
[$n_1$,$p_1$($p_2$)] labels, where $n_1$ is the number of neutrons in the 
intruder $N=4$ orbitals, and $p_1$ and $p_2$ are the number of protons
in the $N=4$ intruder and $N=5$ hyperintruder orbitals. The $p_2$ number is 
omitted when respective orbitals are not occupied. 
}
\label{E-ERLD}
\end{figure}

\begin{figure}[ht]
\centering
\includegraphics[width=8.8cm]{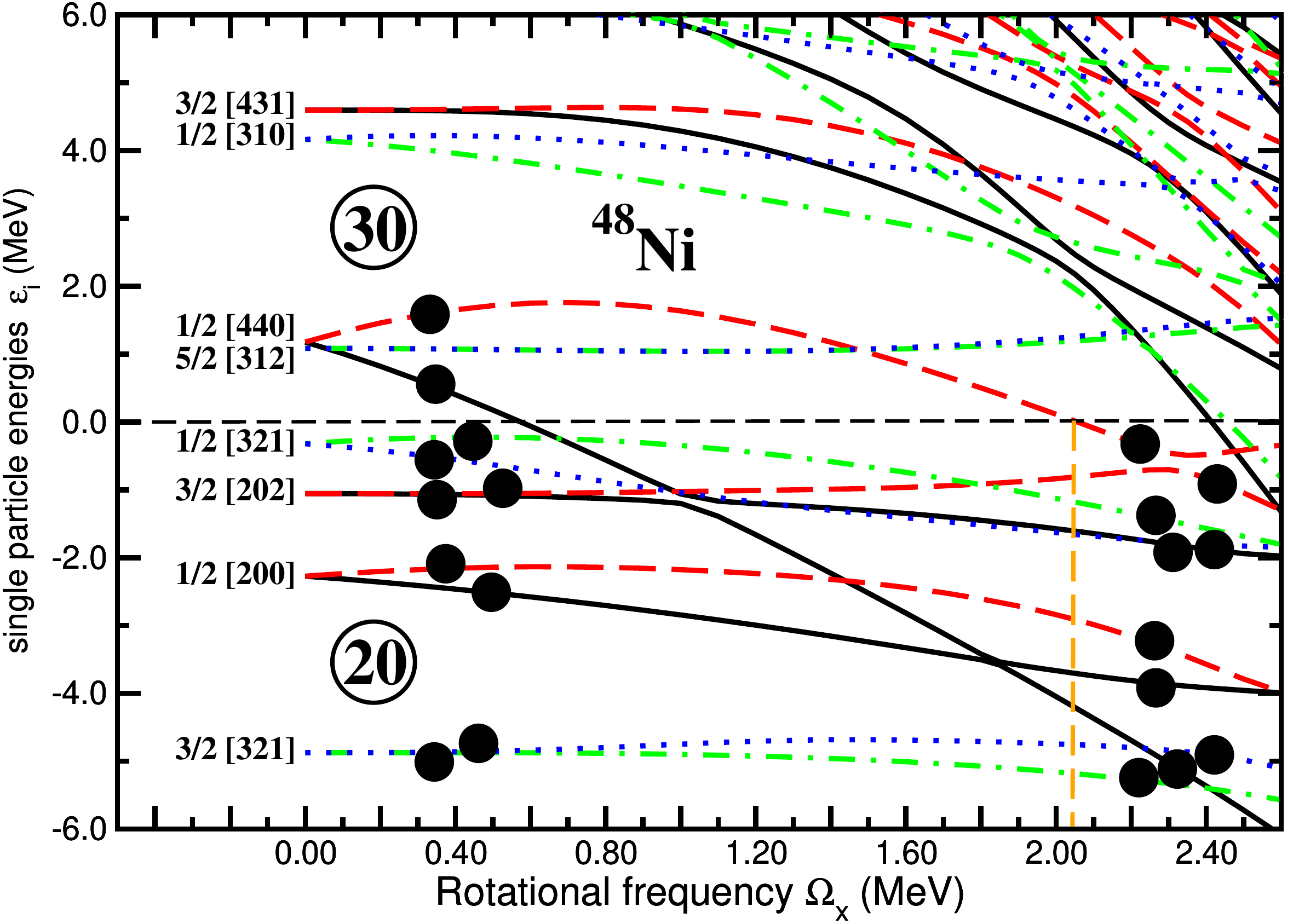}
\caption{The same as Fig.\ \ref{routh-Mg46} but for proton routhians in the
[0,2] configuration of $^{48}$Ni.}
\label{routh-Ni48}
\end{figure}

The birth of particle-bound rotational bands provides a  mechanism for the extension
of nuclear landscape  to neutron numbers which are larger than those of the neutron drip
line in non-rotating nuclei (Ref.\ \cite{AIR.19}). For example, at spin zero $^{46}$Mg is the 
last bound even-even nucleus \cite{AARR.14}. However, neutron bound rotational states are 
predicted at  non-zero spins also in $^{48,50}$Mg which are unbound at spin zero (see Figs.\ 
6(i) and (j) in Ref.\ \cite{AIR.19}).  Thus, rotational excitations allow to extend the nuclear 
landscape for  the Mg isotopes by four neutrons.

On-going investigation \cite{TA.20} reveals that similar mechanisms are active also in 
the nuclei in the vicinity of proton drip line.  However, in this case the birth of particle (proton)
bound rotational bands and the extension of nuclear landscape beyond spin zero 
proton drip line\footnote{See Refs.\ \cite{AARR.13,AARR.14,Eet.12} for the state-of-the art 
predictions of the  position of proton drip line.}  emerge from proton intruder orbitals 
initially located at positive  energies which with increasing rotational 
frequency dive into negative energy domain because of high-$j$ angular momentum 
content leading to large Coriolis force.  However, because of the presence of the Coulomb 
barrier, the part of the rotational band with at least one occupied proton single-particle state 
having positive energy will have discrete rotational states which can decay also by proton
emission. This is contrary to the situation near neutron drip line where occupied neutron state(s) with 
positive energy result in resonance part of the band with rotational states having 
finite width.  The nucleonic configuration will be proton bound for the case of negative 
energy of  highest occupied proton orbital. Similar to very neutron-rich nuclei, the dive of 
intruder proton orbitals into nucleonic potential can trigger the transition from proton 
emitting part of rotational band at low spin to proton bound one at higher spin.

   These features are illustrated in Figs.\ \ref{E-ERLD} and \ref{routh-Ni48}. The $^{50}$Ni 
and $^{48}$Ni are the last proton bound and the first proton unbound Ni nuclei at spin 
zero \cite{AARR.14}.  At $I=0$, these nuclei are spherical because of the presence
of the $Z=28$ and $N=20$ shell closures. However, particle-hole excitations across these
gaps lead to the development of large and extreme deformations (see Ref.\ \cite{AR.16}) as 
well as to the occupation of positive energy proton orbitals in excited configurations.
As a consequence, these bands are proton emitting. However, some configurations
built on intruder proton orbitals, such as [0,2] and [2,42] in $^{50}$Ni and [0,2] in $^{48}$Ni, 
which are proton emitting at low and moderate spins become proton bound at high spins. 
The underlying microscopic mechanism is illustrated in Fig.\ \ref{routh-Ni48}.  Proton 
$1/2[440](i=\pm)$ intruder orbitals, occupied in the [0,2] configuration of $^{48}$Ni, are 
located at positive energy at low spins. However, both of them dive into negative energy 
domain with increasing rotational frequency so that the [0,2] configuration become particle 
bound  at $\Omega_x=2.04$ MeV ($I\approx 26\hbar$).

\section{Conclusions}
\label{Summary}

  In conclusion, it was illustrated that traditional limits of nuclear landscape 
can be substantially expanded when considering the  mechanisms alternative
to those discussed before. In the region of hyperheavy nuclei, the transition 
from ellipsoid-like nuclear shapes to toroidal ones provides a substantial 
increase of nuclear landscape. Rotational excitations
provide an alternative mechanism of the extension of nuclear landscape beyond
the limits defined at spin zero. In both cases, the collective coordinates 
related to nuclear shapes play an important role in extending nuclear 
landscape. In hyperheavy nuclei, they drive the nuclear systems from ellipsoidal-like
to toroidal shapes. In rotating nuclei, they transform the system from spherical
or normal deformed ground states to extremely elongated (super-, hyper- 
and megadeformed) shapes at high spins. This creates favorable positions of 
intruder  orbitals with respect of zero energy threshold which combined with the 
action of collective  rotation allows to extend the limits of nuclear landscape beyond
spin zero ones. 
\\
\\
This material is based upon work supported by the U.S. Department of Energy,
Office of Science, Office of Nuclear Physics under Award No. DE-SC0013037 
and U.S. Department of Energy, National Security Administration under Award 
No. DE-NA0002925.

\end{document}